\documentclass[sigconf]{acmart}

\AtBeginDocument{%
  }

\acmConference[TREC 2025]{TREC 2025 BioGen Track}{December 11--12, 2025}{Gaithersburg, MD, USA}

\setcopyright{none}
\renewcommand\footnotetextcopyrightpermission[1]{}
\usepackage{xcolor}
\usepackage{enumitem}
\usepackage[most]{tcolorbox}
\usepackage{graphicx}
\usepackage{multirow}
\usepackage{amsmath}
\usepackage{url}
\usepackage{booktabs}

\definecolor{gepurple}{RGB}{108, 29, 165}
\definecolor{geback}{RGB}{250, 248, 255}

\begin{document}

\title[Contradiction-Aware Biomedical QA]{Negation is Not Semantic: Diagnosing Dense Retrieval Failure Modes for Trade-offs in Contradiction-Aware Biomedical QA}

\author{Soumya Ranjan Sahoo}
\affiliation{%
  \institution{GE HealthCare}
  \city{Bangalore}
  \country{India}}
\email{Soumya.Sahoo1@gehealthcare.com}

\author{Gagan N.}
\affiliation{%
  \institution{GE HealthCare}
  \city{Bangalore}
  \country{India}}
\email{Gagan.N@gehealthcare.com}

\author{Sanand Sasidharan}
\affiliation{%
  \institution{GE HealthCare}
  \city{Bangalore}
  \country{India}
}

\author{Divya Bharti}
\affiliation{%
  \institution{GE HealthCare}
  \city{Bangalore}
  \country{India}
}

\renewcommand{\shortauthors}{Sahoo et al.}

\begin{abstract}
Large Language Models (LLMs) have demonstrated remarkable capabilities in biomedical question-answering tasks, yet their tendency to generate plausible yet unverified claims poses significant risks in clinical contexts. To mitigate the clinical risks of LLM hallucinations, the TREC 2025 BioGen track mandates grounded answers that explicitly surface contradictory evidence (Task A) and the generation of narrative-driven, fully attributed responses (Task B). Addressing the critical absence of target ground truth, we present a proxy-based development framework utilizing the SciFact dataset to systematically optimize retrieval architectures. Our iterative evaluation revealed a ``Simplicity Paradox'': complex adversarial dense retrieval strategies failed catastrophically on contradiction detection (MRR 0.023) due to Semantic Collapse, where negation signals were indistinguishable in vector space. Furthermore, we identified a distinct Retrieval Asymmetry: while filtering dense embeddings improved contradiction detection, it degraded support recall, compromising holistic reliability. We resolve this via a Decoupled Lexical Architecture utilizing a unified BM25 backbone to balance semantic support recall (0.810) with precise contradiction surfacing (0.750). This approach achieves the highest Weighted MRR (0.790) on the proxy benchmark while remaining the only computationally viable strategy for scaling to the 30-million-document PubMed corpus. For answer generation, we introduce Narrative-Aware Reranking and One-Shot In-Context Learning, which improved citation coverage from 50\% (zero-shot) to 100\%. Our official TREC evaluation results confirm these findings: our system ranks 2nd among all teams on Task A contradiction F1 and 3rd out of 50 runs on Task B citation coverage (98.77\%), achieving zero citation contradict rate. Our work transforms LLMs from stochastic generators into honest evidence synthesizers, demonstrating that epistemic integrity in biomedical AI requires prioritizing lexical precision and architectural scalability over isolated metric optimization.
\end{abstract}


\maketitle

\section{Introduction}

Large language models (LLMs) are increasingly applied to biomedical question answering, where clinicians and researchers expect concise, trustworthy, and verifiable outputs. Yet even state-of-the-art systems can produce confident but unsupported claims, a risk that undermines trust in medical AI. Recent work shows that retrieval-augmented generation (RAG) can reduce hallucination but still struggles with factual consistency, evidence alignment, and contradiction handling \citep{xiong2024mirage,he2024covrag}. In biomedical contexts, this risk is magnified: a system that presents only supportive findings while omitting contradictory results can reinforce confirmation bias and lead to unsafe interpretations \citep{mao2024rafe,bmretriever2024,petroni2024irragsigirtoc}.

To address these shortcomings, the TREC 2025 BioGen track explicitly evaluates contradiction-aware grounding and per-sentence citation attribution. Systems must synthesize information from the PubMed corpus while surfacing dissenting or uncertain findings alongside supportive ones. However, the absence of ground-truth labels in the target corpus makes systematic development and evaluation difficult. Most retrieval pipelines still optimize for topical similarity rather than logical entailment, leading models to conflate ``about the same topic'' with ``agrees with the claim.'' Contradictory evidence is rarer and more linguistically nuanced, often expressed through negation, hedging, or population-specific conditions, thus making it hard for generic dense retrieval to capture.

In this work, we adopt \textbf{epistemic integrity} as our central design principle. A biomedical QA system should act as an honest evidence synthesizer that reveals uncertainty instead of masking it. Supporting and contradictory evidence follow different retrieval dynamics and therefore require independent optimization. We operationalize this stance through a proxy-based development framework and two task-specific architectures aligned to BioGen's dual objectives.

To enable reproducible development despite the lack of labeled data, we employ the SciFact dataset as a \textbf{proxy environment} for PubMed. SciFact provides claim-level annotations for support and contradiction, allowing systematic benchmarking of retrieval, reranking, and contradiction-detection components before transferring the optimized design to the large-scale BioGen corpus \citep{wadden2020scifact}.

To improve factual grounding, our framework \textbf{decouples supporting and contradictory pipelines}. The support pipeline emphasizes semantic precision using BM25 retrieval and cross-encoder reranking. In contrast, the contradiction pipeline prioritizes high recall through expanded lexical retrieval, negation-aware filtering, and calibrated natural-language inference. This decoupling mitigates the \textit{``entailment trap,''} where dense retrievers over-rank passages that agree with a claim while missing explicit refutations. For example, one study might report a biomarker as correlated with disease, while another finds ``no association after adjustment for confounders.'' Without negation-sensitive retrieval, standard systems typically over-rank the first result and ignore the second.

The BioGen challenge introduces \textbf{two complementary tasks} that reflect different reasoning stages. Task A (Grounding) involves \textit{post-hoc verification}: linking a fixed answer sentence to new supporting and contradictory PubMed IDs while prioritizing contradictions when both exist. Task B (Attribution) requires generating an answer with inline citations for every sentence under a strict 250-word limit. Although related, these tasks demand distinct architectures. Task A follows a generate-then-retrieve approach, while Task B employs retrieve-then-generate reasoning. Treating them separately enables clearer optimization and analysis.

We study four questions that drive our design and evaluation:
\begin{enumerate}
    \item \textbf{Proxy transfer:} How effectively can a proxy-labeled corpus such as SciFact guide architecture and component choices for contradiction-aware QA on unlabeled biomedical corpora?
    \item \textbf{Separation principle:} What degree of architectural separation between support and contradiction pipelines (retrieval depth, reranking, negation filtering, NLI calibration) is necessary to maximize contradiction recall without harming support precision?
    \item \textbf{Attribution discipline:} In generative settings, how does narrative-aware reranking coupled with one-shot in-context learning affect sentence-level citation \emph{coverage}, \emph{density}, and \emph{faithfulness} under tight word limits?
    \item \textbf{Failure modes at scale:} Which recurrent failure modes --- entailment misclassification, retrieval--reranking interference, calibration drift --- emerge across tasks, and which design patterns mitigate them under PubMed-scale constraints?
\end{enumerate}

This study advances the field of contradiction-aware biomedical question answering through the following key contributions: (1) a proxy-based evaluation protocol for contradiction-aware biomedical QA without target labels; (2) a decoupled retrieval architecture that separates high-precision support retrieval from high-recall contradiction detection using negation-aware filtering and calibrated NLI; (3) a narrative-driven RAG pipeline that enforces per-sentence citation through one-shot in-context learning; and (4) an ablation across multiple variants revealing failure modes and quantifying the trade-off between statistical optimization and epistemic balance.

Our results demonstrate that this design achieves measurable improvements in contradiction retrieval while maintaining strong support accuracy. On the SciFact proxy, the decoupled architecture increases contradiction recall compared to dense-only systems that capture topical but not logical similarity. For generative attribution, narrative-aware reranking and one-shot prompting achieve complete sentence-level citation coverage and higher citation density without compromising fluency. Together, these findings show that epistemic integrity --- balancing supportive and contradictory evidence --- can be operationalized through principled architectural decoupling and proxy evaluation.

Finally, this paper focuses on \textbf{scalable and transparent} solutions that surface contradictions and maintain clinical relevance under realistic computational limits. We discuss limitations of proxy transfer, temporal sensitivity, and model dependence, and outline future directions such as temporal evidence weighting, improved negation modeling, and open-source deployment. The remainder of the paper reviews related work, formalizes the BioGen tasks, details our architectures, presents experimental results, and concludes with limitations and future research paths.

\section{Related Work}

\subsection{Evolution of Biomedical Generative Retrieval}

The TREC BioGen 2024 track introduced a significant shift in biomedical question answering (QA) by treating it as a joint retrieval-and-generation problem. This benchmark aimed to address a major weakness of Large Language Models (LLMs): their tendency to generate fluent but unsupported claims. By requiring systems to attribute each statement to specific supporting documents, the track moved the evaluation focus from simple fluency to verified evidence grounding. Most participating teams used multi-stage pipelines, combining standard retrieval with LLM-based synthesis, which reflects the difficulty of maintaining factual accuracy using generative approaches alone \cite{Gupta2024OverviewOT}.

TREC BioGen 2025 extends this scope significantly by splitting the problem into two distinct objectives. \textbf{Task A (Grounding)} elevates contradiction detection from a secondary feature to a primary objective, requiring systems to identify citations that explicitly refute existing claims. \textbf{Task B (Attribution)} strengthens the generative aspect by requiring detailed sentence-level citations. Methodologically, the 2025 track maintains continuity with a stable PubMed baseline but shifts the focus from ``\textit{Can you cite?}'' to ``\textit{Can you systematically ground, support, and challenge a biomedical answer?}''

\subsection{Reference Attribution vs. Contextual Grounding}

Although they are often used interchangeably, reference attribution and contextual grounding operate at different levels.

\textbf{Reference Attribution} is a fine-grained requirement where each factual statement in a generated answer must be explicitly linked to one or more source documents. In the BioGen framework, this is measured by penalizing citations that do not logically support the associated sentence. This makes attribution directly auditable; a clinician can easily verify if a specific claim appears in the cited evidence.

\textbf{Contextual Grounding}, on the other hand, is a broader requirement that the model must base its outputs on a given context and avoid hallucinations. Retrieval-Augmented Generation (RAG) acts as a bridge between these concepts. While standard RAG ensures grounding by using retrieved text, it does not guarantee strong reference attribution, as models often combine information from multiple passages without marking the exact source. BioGen-style systems extend RAG by enforcing an explicit source-tracking mechanism, thereby transforming the generator from a creative engine into a verifiable synthesizer.

\subsection{The Challenge of Contradiction Detection}

Historically, contradiction detection has been treated as a subtask of Natural Language Inference (NLI). However, recent studies indicate that this task is much harder in real-world retrieval than what benchmark scores suggest.

Early benchmarks like SNLI and MultiNLI unintentionally encouraged models to learn shallow shortcuts such as matching similar words or finding negation words like ``no'' instead of true reasoning. Renjit et al.~\cite{renjit2024study} explicitly flag these ``dataset artifacts'' as major obstacles for reliable contradiction detection.

This observation is critical for interpreting modern failures. McCoy et al.~\cite{mccoy2019right} demonstrated that BERT models often fail on adversarial datasets, systematically misclassifying contradictions as agreements when there is high word overlap. This finding supports the \textbf{Semantic Collapse} observed in our dense retrieval experiments, where negation signals get lost because the topics are very similar.

Furthermore, Laban et al.~\cite{laban2022summac} reported that standard NLI models perform poorly when detecting inconsistencies between long documents and summaries. More recently, Luo et al.~\cite{luo2024factual} have shown that even advanced LLM-based metrics fail to detect subtle contradictions and are overly influenced by surface similarity. Collectively, these works emphasize that contradiction detection remains a significant challenge where semantic models often default to keyword matching, a limitation our architecture addresses by using a decoupled lexical retrieval strategy.

\section{Task Formulation}

The TREC 2025 BioGen track introduces two complementary tasks focused on evidence-based biomedical question answering with explicit emphasis on surfacing contradictory evidence. Although Task A (Grounding Answer) and Task B (Reference Attribution) share the common goal of connecting biomedical claims with evidence, they differ fundamentally in their scope, inputs, and system requirements. Understanding these differences is crucial for architecture design and evaluation strategy.

\subsection{Task A: Grounding Answer}

\textbf{Task Objective:} Ground pre-generated answer sentences by identifying both supporting evidence (beyond existing citations) and contradictory evidence that challenges the claim.

\begin{description}
    \item[Input:] The task takes the following inputs:
    \begin{itemize}
        \item Biomedical question $q$
        \item Answer sentence $s$
        \item Existing (outdated) supporting PMIDs $P_{old}$
        \item PubMed corpus $\mathcal{C}$
    \end{itemize}

    \item[Output:] The system must return:
    \begin{itemize}
        \item Supporting PMIDs $P_{supp}$ where $|P_{supp}| \leq 3$
        \item Contradicting PMIDs $P_{contra}$ where $|P_{contra}| \leq 3$
    \end{itemize}

    \item[Constraints:]
    \begin{enumerate}
        \item All PMIDs must be selected from the corpus ($P_{supp}, P_{contra} \subset \mathcal{C}$).
        \item $P_{supp}$ must be \textit{additional} to $P_{old}$ (i.e., $P_{supp} \cap P_{old} = \emptyset$).
        \item If both supporting and contradicting evidence are found, priority is given to $P_{contra}$.
    \end{enumerate}
\end{description}

\subsection{Task B: Reference Attribution}

\textbf{Task Objective:} Generate comprehensive, evidence-grounded answers with inline citations for each sentence.

\begin{description}
    \item[Input:] The task takes the following inputs:
    \begin{itemize}
        \item Biomedical question $q$
        \item Topic $\mathcal{T}$ (optional context)
        \item Narrative $\mathcal{N}$ (optional context)
        \item PubMed corpus $\mathcal{C}$
    \end{itemize}

    \item[Output:] The system must generate:
    \begin{itemize}
        \item Generated answer $A = \{s_1, s_2, \dots, s_n\}$
        \item Citation set $C_i$ for each sentence $s_i$, where $|C_i| \leq 3$
    \end{itemize}

    \item[Constraints:]
    \begin{enumerate}
        \item The total answer length must be less than 250 words ($|A| \leq 250$).
        \item All PMIDs in citation sets $C_i$ must be selected from corpus $\mathcal{C}$.
    \end{enumerate}
\end{description}

\subsection{Task Relationship}

Task A focuses on post-hoc verification, that is, grounding fixed answer text with appropriate citations. Task B, however, requires end-to-end answer generation, jointly answering a biomedical question and citing appropriate citations at the sentence level. Although seemingly complementary, both tasks require fundamentally separate architectures. Task A uses post-hoc retrieval (\textit{generate-then-retrieve}) to ground fixed answer text, enabling aggressive retrieval strategies for high-quality evidence retrieval. Task B uses retrieval-augmented generation (\textit{retrieve-then-generate}) where evidence retrieval precedes and influences answer generation. These architectural differences necessitate separate system designs rather than treating Task A as merely a foundation for Task B.

\section{Methodology}

To address the distinct architectural requirements of post-hoc verification (Task A) and generative attribution (Task B), we propose a modular framework. While both tasks share a common retrieval backbone indexed from the target PubMed collection $\mathcal{C}$, we diverge in our optimization strategies: Task A employs a \textit{Decoupled Retrieval-Reranking} architecture to handle class imbalance between support and contradiction, while Task B utilizes a \textit{Narrative-Driven RAG} architecture with One-Shot In-Context Learning.

\begin{figure*}[t]
    \centering
    \IfFileExists{fig/biogen-architecture-mod.png}{%
        \includegraphics[width=\textwidth]{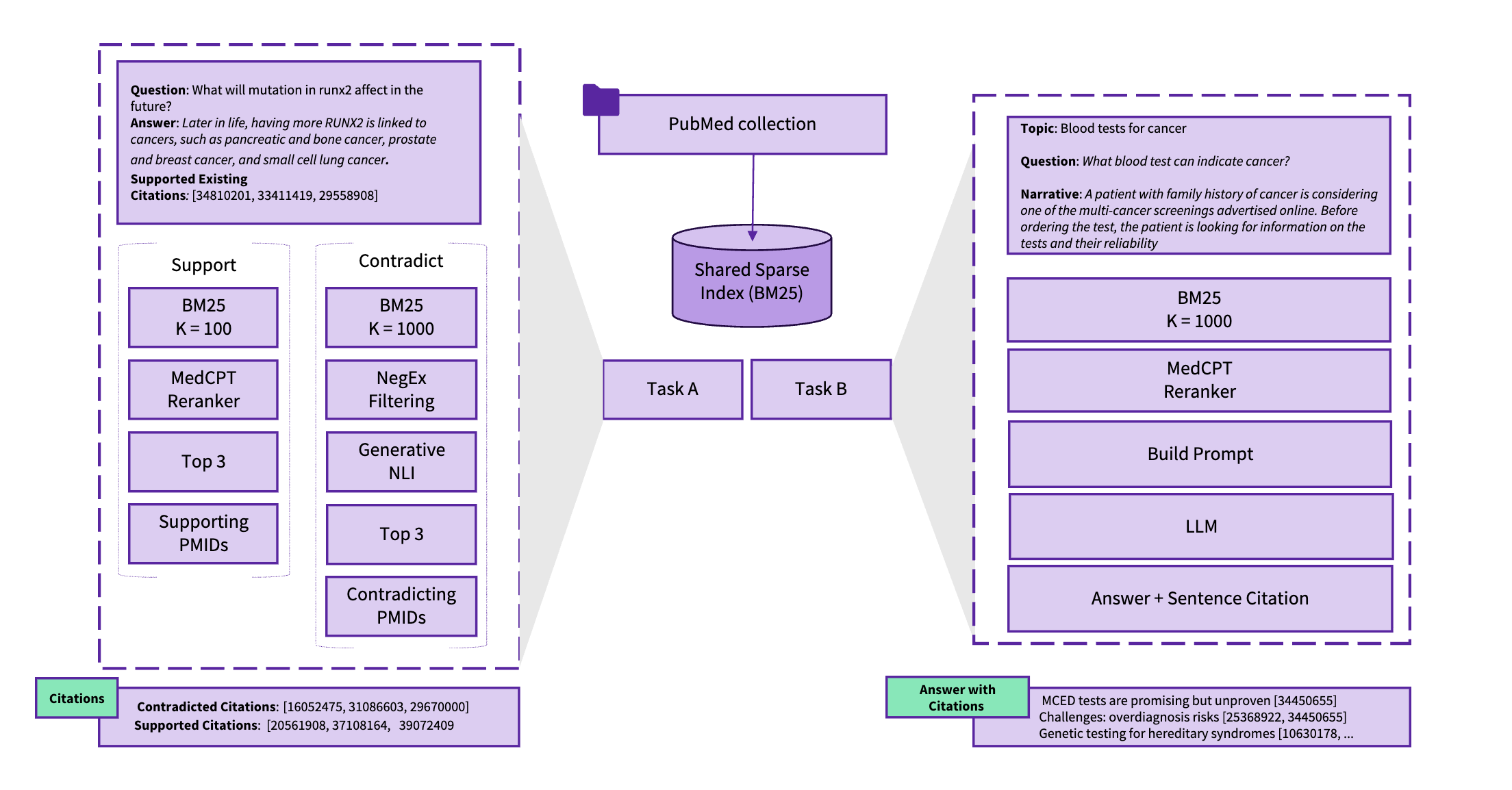}%
    }{%
        \begin{tcolorbox}[colback=gray!5,colframe=black,title=Figure Placeholder]
        The file \texttt{fig/biogen-architecture-mod.png} was not found.
        Replace this box with the system architecture figure before final submission.
        \end{tcolorbox}
    }
    \caption{\textbf{Schematic representation of the proposed system architecture.} The framework integrates a \textbf{Decoupled Retrieval-Reranking} module for Task A to balance support and contradiction detection, and a \textbf{Narrative-Aware RAG} module for Task B utilizing One-Shot In-Context Learning for controlled-citation generation.}
    \label{fig:overall-architecture}
\end{figure*}

\noindent
Figure~\ref{fig:overall-architecture} summarizes the unified retrieval backbone and the task-specific reasoning pipelines used in our system. Both Task~A and Task~B operate over the same sparse BM25 index constructed from the full PubMed collection, but diverge significantly in how retrieved documents are processed. Task~A performs post-hoc grounding of a fixed answer sentence using two independent pipelines optimized for different epistemic functions: a high-precision support branch retrieves a shallow ($K=100$) candidate set that is reranked with MedCPT, while a high-recall contradiction branch explores a deeper ($K=1000$) lexical pool and applies NegEx-based negation filtering followed by sentence-level generative NLI. These two paths surface up to three supporting and three contradicting PMIDs, enabling explicit exposure of evidence that reinforces or refutes the claim.

In contrast, Task~B performs retrieve-then-generate reasoning: the question and narrative query the same BM25 index at a deeper retrieval depth ($K=1000$), MedCPT reranks the results, and the top abstracts are converted into a structured prompt. This prompt is passed to the LLM, which synthesizes a $<250$-word answer in which every sentence must carry one to three citations. The diagram thus highlights the central architectural principle of our system --- while retrieval is shared and lexical, the downstream reasoning branches are deliberately decoupled to satisfy the distinct epistemic requirements of grounding (Task~A) and attribution (Task~B).

\subsection{Task A Architecture: Decoupled Evidence Grounding}

The core challenge in Task A is the ``Entailment Trap'' where dense retrievers confuse \textit{topical similarity} (discussing the same entities) with \textit{logical entailment} (confirming the claim) \cite{Hossain2020}. Furthermore, contradictory evidence is statistically rare and linguistically distinct from supporting evidence. To mitigate these issues, we decouple the retrieval systems into two separate pipelines.

\subsubsection{The Support Pipeline ($\mathcal{P}_{supp}$)}

This pipeline optimizes for semantic alignment to populate $P_{supp}$.

\begin{enumerate}
    \item \textbf{Initial Retrieval:} We query $\mathcal{C}$ using the Pyserini implementation of BM25 \citep{Lin2021} with the concatenated query $q \oplus s$, retrieving a candidate set $C_{supp}$ ($k=100$). We utilize default parameters ($k_1=0.9, b=0.4$).
    \item \textbf{Exclusion Filter:} We enforce strict novelty by filtering out any document $d \in C_{supp}$ if $d \in P_{old}$.
    \item \textbf{Semantic Reranking:} We employ the \textbf{MedCPT Cross-Encoder} \citep{Jin2023}, a model pre-trained on 255 million user clicks from PubMed logs. We utilize the \texttt{ncbi/medcpt-cross-encoder} checkpoint to score pairs $(s, d)$.
    \item \textbf{Selection:} We select the top-3 documents maximizing the cross-encoder score as $P_{supp}$.
\end{enumerate}

\subsubsection{The Contradiction Pipeline ($\mathcal{P}_{contra}$)}

This pipeline prioritizes high recall and negation sensitivity to populate $P_{contra}$.

\begin{enumerate}
    \item \textbf{High-Recall Retrieval:} Recognizing that contradictions are rare, we expand retrieval to $k=1000$ using BM25. Lexical retrieval is preferred here as negation markers (e.g., ``absence of'', ``no evidence of'', ``no signs of'') are often captured better by exact matching than by dense embeddings \citep{Thakur2021}.
    \item \textbf{Sentence Segmentation \& Negative Filtering:} We decompose documents into sentences and apply a rule-based filter $\mathcal{F}_{neg}$ using 23 common clinical negation patterns derived from \textbf{NegEx} \citep{Chapman2001}. Only sentences containing explicit negation cues are retained.
    \item \textbf{Generative Classification:} Surviving sentences are processed by a \textbf{MedNLI-tuned T5-Base} model \citep{Romanov2018}. Unlike probabilistic classifiers (e.g., BERT-CLS heads), we employ a generative approach where the model must explicitly decode the token ``contradiction.''
    \item \textbf{Selection:} The first 3 instances classified as contradictions are selected as $P_{contra}$. Selecting the \textit{first} 3 matches (based on BM25 rank order) is a heuristic assumption that BM25 rank correlates with relevance.
\end{enumerate}

\subsection{Task B Architecture: Narrative-Driven RAG}

For Task B, we treat the tuple $(q, \text{Narrative}, \text{Topic})$ as the input context. Our approach introduces two key innovations: Narrative-Aware Reranking and One-Shot In-Context Learning \citep{Brown2020}.

\subsubsection{Narrative-Aware Reranking}

Standard RAG systems \citep{Lewis2020} typically retrieve evidence using only the question $q$. We hypothesize that the provided \textit{Narrative} --- which details the user's background intent and exclusion criteria --- offers a superior semantic representation for evidence selection. To optimize this, we employ a \textbf{Decoupled Query Formulation} strategy:

\begin{itemize}
    \item \textbf{Stage 1 (Broad Retrieval):} We retrieve an initial pool $C_{init}$ ($k=1000$) using BM25 exclusively on the question $q$. Note that the choice of using sparse retrieval in BM25 was an informed decision from our experiments in solving Task A.
    \item \textbf{Stage 2 (Contextual Reranking):} We rerank $C_{init}$ using the MedCPT Cross-Encoder. Crucially, we switch the query input to the \textit{Narrative} text.
    \item \textbf{Selection:} The top-10 documents ($D_{top10}$) are selected as the context window for generation.
\end{itemize}

\paragraph{Rationale for Input Decoupling:}
We deliberately exclude the \textit{Topic} field during the retrieval and reranking phases to optimize the signal-to-noise ratio. First, in lexical retrieval (Stage 1), \textit{Topic} terms are often broad categorical labels (e.g., ``Iron levels in Covid-19'') which, if concatenated with $q$, induce query drift by prioritizing documents that saturate the topic keywords rather than addressing the specific clinical inquiry. Second, in dense reranking (Stage 2), the \textit{Narrative} field semantically subsumes the \textit{Topic} while providing higher granularity. Excluding the redundant \textit{Topic} string minimizes token consumption, maximizing the context window available for candidate document tokens within the Cross-Encoder's strict limit (512 tokens). The \textit{Topic} is reintroduced only during the final generation stage to prime the LLM's global context.

\subsubsection{One-Shot In-Context Learning}

To ensure robust attribution without parameter updates, we employ In-Context Learning (ICL). We construct a prompt $\Psi$ incorporating a rigorous set of constraints and a single ``Gold Standard'' exemplar.
\begin{equation}
 \Psi = I_{constraints} \oplus E_{shot} \oplus D_{top10} \oplus (q, \text{Narrative}, \text{Topic})
\end{equation}
\begin{itemize}
    \item \textbf{$I_{constraints}$:} Specifies strict rules: $|A| < 250$ words, journalistic tone, and mandatory citation indices $[i]$ for every sentence.
    \item \textbf{$E_{shot}$:} A complete example (Topic: Iron/Ferritin) demonstrating the required narrative flow (Context $\to$ Mechanism $\to$ Significance) and citation density. This leverages the ability of LLMs to mimic the structural properties of demonstrations \citep{Min2022}.
\end{itemize}
We utilize \textbf{GPT-4o} (checkpoint \texttt{2024-11-20}) to generate the final answer set $A$ and citation sets $C_i$.

\section{Experiments and Analysis}

\subsection{Task A: Proxy-Based Grounding Evaluation}

\subsubsection{SciFact as Proxy Environment}

To systematically develop our retrieval system without ground truth labels or relevance judgements on the target corpus, we employed the SciFact dataset \cite{wadden-etal-2020-scifact} as a proxy development environment. SciFact contains 5,183 PubMed abstracts labeled with support/contradiction annotations, sharing strong similarities with TREC BioGen in domain and task structure. We evaluated our systems on a subset of the development set ($N_{total}=188$, where $N_{supp}=124$ and $N_{contra}=64$).

\subsubsection{Granularity Analysis: Sentence vs. Document NLI}

A critical architectural decision across all our experimental variants was the adoption of sentence-level classification rather than document-level inference. Prior work has shown that document-level entailment models suffer from context dilution in factual texts. Our preliminary validation confirmed this: sentence-level decomposition provided a $4.8\times$ improvement in contradiction detection compared to document-level inference. Consequently, all subsequent system variants employ sentence-level processing.

\subsubsection{System Variants}

We systematically evolved the pipeline through five architectural paradigms, categorized by their retrieval depth and pipeline independence.

\paragraph{Group A: Single-Pipeline Retrievers ($k=500$)}
These variants utilize a uniform retrieval strategy for both support and contradiction tasks, fetching a static pool of 500 candidates per query.
\begin{itemize}
    \item \textbf{Variant 1: Naive BM25 Baseline.} A purely lexical approach using BM25 on the combined query ($q+a$). It relies entirely on the MedNLI model for classification without any heuristic filtering. This serves as the baseline to measure the impact of explicit negation gates.
    \item \textbf{Variant 2: Hybrid Semantic Retrieval.} Integrates dense vector representations (Snowflake Arctic-Embed-M-v2.0 \cite{snowflake2024arctic}) fused with BM25 using Reciprocal Rank Fusion (RRF) \cite{Cormack2009ReciprocalRF}.
\end{itemize}

\paragraph{Group B: Decoupled \& Multi-Stage Retrievers}
Recognizing the asymmetric nature of support (precision-oriented) versus contradiction (recall-oriented) retrieval, these variants decouple the pipelines either through \textit{retrieval methodology} or \textit{retrieval depth}.
\begin{itemize}
    \item \textbf{Variant 3: Decoupled Dense + Filter.} Uses pure dense retrieval with asymmetric depths ($k=100$ for support, $k=1000$ for contradiction) and applies NegEx rule-based filtering to contradiction candidates. Tests whether dense embeddings, when aided by filters, can outperform BM25.
    \item \textbf{Variant 4: Multi-Query Adversarial Fusion.} Decouples via retrieval \textit{strategy}: generates 25 negation-heavy query variations (e.g., ``$q$ refutes $a$'', ``no effect of $a$'', ``fails to show'') from templates, each retrieving $k=200$ via dense search. Results are fused via RRF into a candidate pool capped at 1200 documents. Candidates were ranked using a 2-way NLI scoring and a custom probabilistic penalty function:
    \begin{equation}
    S(d) = P(\text{Con}\mid d) - 0.5 \cdot P(\text{Ent}\mid d) + \text{CueBonus}
    \end{equation}
    This tests if mathematical probability calibration can outperform rule-based filtering (details in Appendix~\ref{app:adversarial}).
    \item \textbf{Variant 5: Decoupled BM25 + Filter (Final).} Our optimized architecture. Decouples via retrieval \textit{depth}: Support uses BM25 ($k=100$) + MedCPT reranking; Contradiction uses BM25 ($k=1000$) + NegEx filter + sentence-level NLI. Tests the hypothesis that lexical retrieval provides better candidate pools for negation detection than dense vectors.
\end{itemize}

\subsubsection{Quantitative Results}

We utilize Weighted MRR as the decisive metric to compare architectures. Given the dataset distribution ($N_{supp}=124, N_{contra}=64$), the weighted score reflects the system's holistic reliability:
\begin{equation}
\text{Weighted MRR} = \frac{N_{supp} \cdot \text{MRR}_{supp} + N_{contra} \cdot \text{MRR}_{contra}}{N_{total}}
\end{equation}
Table~\ref{tab:scifact_results} presents the comparative performance.

\begin{table}[h]
\centering
\caption{Performance of grounding architectures on the SciFact proxy. Variant 5 achieves the highest Weighted MRR (0.790). Note the failure of V4 on contradictions despite high support recall, and the holistic superiority of V5 over the dense-based V3.}
\label{tab:scifact_results}
\resizebox{\columnwidth}{!}{%
\begin{tabular}{lcccc}
\toprule
\textbf{System Variant} & \textbf{MRR (Sup)} & \textbf{MRR (Con)} & \textbf{Weighted MRR} & \textbf{Rank} \\
\midrule
\textit{Group A: Single-Pipeline} & & & & \\
Var 1: Naive BM25 & 0.927 & 0.109 & 0.649 & 5 \\
Var 2: Hybrid RRF & 0.859 & 0.385 & 0.698 & 3 \\
\midrule
\textit{Group B: Decoupled} & & & & \\
Var 4: Adversarial & 0.988 & 0.023 & 0.660 & 4 \\
Var 3: Dense + Filter & 0.786 & 0.766 & 0.779 & 2 \\
\textbf{Var 5: BM25 + Filter} & 0.810 & \textbf{0.750} & \textbf{0.790} & \textbf{1} \\
\bottomrule
\end{tabular}%
}
\end{table}

\subsubsection{Comparative Analysis}

\paragraph{The Failure of Complexity (V4 vs.\ V1)}
While the naive V1 baseline failed on contradictions (0.109) due to lack of filtering, the ``over-engineered'' V4 (Adversarial) performed even worse (0.023) despite its massive retrieval pool (limit 1200). The complex penalty formula likely caused calibration issues, suppressing relevant contradictions that had mixed NLI signals. This validates the principle that in zero-shot biomedical verification, heuristic simplicity (binary filtering) often outperforms probabilistic complexity.

\paragraph{Dense vs.\ Lexical Backbones (V3 vs.\ V5)}
The comparison between V3 and V5 is critical, as both utilize the effective NegEx filter strategy.
\begin{itemize}
    \item \textbf{Variant 3 (Dense)} achieved a strong contradiction score (0.766). However, applying the strict NegEx filter to dense candidates degraded its support performance compared to baselines (0.786). This suggests that dense retrievers rely on semantic ``smoothness''; enforcing strict keyword constraints disrupts their ranking logic for supporting evidence.
    \item \textbf{Variant 5 (BM25)} achieved the best overall balance (Weighted MRR: 0.790). While its contradiction score (0.750) is comparable to V3, its support performance is more robust (0.810).
\end{itemize}

\paragraph{Selection of Final Architecture}
We selected Variant 5 not only for its highest Weighted MRR, but also for \textbf{computational scalability}. Scaling the dense index of V3 (and especially the multi-query inference of V4) to the 30 million documents of the BioGen corpus would require prohibitive memory and compute resources. Variant 5 delivers superior holistic accuracy using a standard, lightweight inverted index, fulfilling both the epistemic and engineering requirements of the task.

\subsection{Task B: RAG Pipeline Optimization}

The retrieve-and-answer generation pipeline evolved through four distinct architectural paradigms to balance the need for clear, flowing narratives with the requirement of properly citing every claim. We analyzed the following variants:
\begin{itemize}
    \item \textbf{V1 (Baseline):} Utilized the Llama-2-7B model with MS-MARCO reranking. This served as a foundational baseline to assess the capabilities of open-source local models against proprietary APIs.
    \item \textbf{V2 (Zero-shot RAG):} Integrated the MedCPT Cross-Encoder with GPT-4o in a zero-shot setting. While retrieval recall was expanded ($k=1000$), this variant relied solely on the question string ($q$) for reranking.
    \item \textbf{V3 (Fallback Ablation):} An intermediate study wherein the general-domain MS-MARCO reranker was reintroduced to verify the necessity of domain-specific biomedical reranking.
    \item \textbf{V4 (Narrative One-shot):} The final architecture which introduced \textit{Narrative-Aware Reranking} (using user intent for selection) and \textit{One-Shot In-Context Learning} to enforce strict citation density.
\end{itemize}

Table~\ref{tab:task_b_metrics} presents the performance comparison of these variants across 30 test topics provided by the organizers.

\begin{table}[h]
\centering
\caption{Performance metrics of generative system variants. \textbf{Coverage} denotes the percentage of sentences containing at least one citation.}
\label{tab:task_b_metrics}
\resizebox{\columnwidth}{!}{%
\begin{tabular}{lcccc}
\toprule
\textbf{Metric} & \textbf{V1} & \textbf{V2} & \textbf{V3} & \textbf{V4} \\
\midrule
Avg Sents per Topic & 3.43 & 7.90 & 6.90 & \textbf{8.10} \\
Citation Coverage & 100\% & 50.6\% & 47.3\% & \textbf{100\%} \\
Avg Citations/Sent & 1.41 & 1.67 & 1.71 & \textbf{1.79} \\
Unsupported Claims & Low & High & High & \textbf{Zero} \\
\bottomrule
\end{tabular}%
}
\end{table}

\subsubsection{Comparative Analysis}

It is observed from the data that \textbf{Variant 4 (Final Submission)} demonstrates objective superiority across all dimensions.

\paragraph{The Attribution-Fluency Trade-off}
Variant 1 (Llama-2) achieved perfect citation coverage (100\%) but produced shallow responses with limited clinical detail (average 3.43 sentences). In contrast, Variants 2 and 3, which employed GPT-4o without examples, generated well-written narratives but exhibited a severe ``attribution gap'' with approximately 50\% of sentences lacking citations, effectively creating unsupported claims that violated the task's fundamental grounding requirement.

\paragraph{Addressing These Limitations in Variant 4}
Our final architecture resolved these issues through two key modifications:
\begin{enumerate}
    \item \textbf{Narrative-Driven Reranking:} By using the user's \textit{Narrative} instead of the question $q$ for cross-encoder reranking, the system better captured implicit user needs, producing more comprehensive answers (8.1 sentences on average).
    \item \textbf{One-Shot In-Context Learning:} Including a detailed exemplar (Iron/Ferritin case) explicitly demonstrated the expected \textit{Context $\to$ Mechanism $\to$ Significance} structure. This successfully enforced consistent citation discipline, achieving the highest citation density (1.79 citations per sentence) while maintaining narrative fluency.
\end{enumerate}

The qualitative impact of these architectural modifications is exemplified in Table~\ref{tab:qualitative_case_study}, which contrasts the hallucinated or biased outputs of earlier versions with the grounded, neutral generation of Variant 4. To facilitate reproducibility, the complete One-Shot System Prompt employed in our final architecture is provided in Box~\ref{box:final_prompt}.

\begin{figure}[h]
    \centering
    \begin{tcolorbox}[
        colback=gepurple!5,
        colframe=gepurple,
        title=\textbf{One-shot Prompt for Task B (Variant 4)},
        arc=3mm,
        boxrule=1pt
    ]
    \footnotesize
    \textbf{System Role:} You are an expert biomedical science communicator. Your goal is to synthesize findings from multiple PubMed articles into a single, clear, and easy-to-understand narrative for an educated audience. Your response will be rigorously judged by human experts based on the following criteria. Adhere to them without exception.

    \vspace{0.2cm}
    \textbf{\#\#\# Rules}

    \begin{enumerate}[leftmargin=*, nosep, label=\arabic*.]
        \item \textbf{Narrative Flow:} Structure your answer like a story: Context $\to$ Mechanism $\to$ Clinical Significance $\to$ Intervention.
        \item \textbf{Evidence Handling:} STRICTLY ONLY use information from the provided documents. Prioritize high-relevance documents (Score $>0.7$). Explicitly state conflicts (e.g., ``While [1] reports X, [2] finds Y'').
        \item \textbf{Sentence Constraint:} Every sentence MUST have at least 1 and a maximum of 3 citations.
        \item \textbf{Citation Format:} Use numerical citations [1], [2], [1,2].
        \item \textbf{Constraints:} Answer length $<250$ words. Journalistic tone.
    \end{enumerate}

    \vspace{0.2cm}
    \textbf{\#\#\# Example of Desired Output Style}

    \textit{Context:} Topic: iron and ferritin levels in COVID-19\ldots

    \textbf{Example Answer:} During infections, a battle for iron takes place between the human body and the invading viruses [1]. The immune system cells need iron to defend the body [1]\ldots If iron balance is disrupted, ferritin levels are high [3], signaling severe disease [4,5].

    \vspace{0.2cm}
    \textbf{\#\#\# Your Task}

    \textbf{Your Context:}
    \begin{itemize}[leftmargin=*, nosep]
        \item Topic: \{topic\}
        \item Question: \{question\}
        \item Narrative: \{narrative\}
    \end{itemize}
    \textbf{Your Provided Evidence:} [List of Documents\ldots]
    \end{tcolorbox}
    \caption{One-shot system prompt used in Variant 4 (Task B). The prompt enforces narrative structure, evidence-only generation, and mandatory per-sentence citations.}
    \label{box:final_prompt}
\end{figure}

\begin{table*}[h]
    \centering
    \begin{tcolorbox}[
        enhanced,
        colback=geback,
        colframe=gepurple,
        boxrule=1.2pt,
        arc=3pt,
        drop shadow southeast,
        title=\textbf{Case Study: Qualitative Output Comparison (Topic 187)},
        coltitle=white,
        fonttitle=\bfseries\large,
        colbacktitle=gepurple
    ]

    \renewcommand{\arraystretch}{1.6}

    \resizebox{\linewidth}{!}{
        \begin{tabular}{|l|p{12cm}|l|}
        \hline
        \multicolumn{3}{|p{16cm}|}{%
            \textbf{Input Context} \newline
            \small
            \textbf{Topic:} Genetically modified foods \newline
            \textbf{Question:} ``Why are genetically modified foods bad?'' \newline
            \textbf{Narrative:} The patient has heard that genetically modified foods are bad for health. She is concerned that consuming genetically modified foods can cause health problems.
        } \\ \hline \hline

        \textbf{Variant} & \textbf{Generated First Sentence} & \textbf{Critique} \\ \hline

        \textbf{V1 (Llama-2)} &
        \textit{``There are several reasons why genetically modified food is bad [12746139].''} &
        Biased \& Erroneous \\ \hline

        \textbf{V2 (Zero-shot)} &
        \textit{``The claim that genetically modified (GM) foods are inherently bad is not supported by the current body of scientific evidence.''} &
        Uncited Hallucination \\ \hline

        \textbf{V3 (Fallback)} &
        \textit{``Genetically modified foods (GM foods) are not inherently bad, but they are associated with certain concerns regarding allergenicity, gene transfer, and outcrossing.''} &
        Uncited Assertion \\ \hline

        \textbf{V4 (Final)} &
        \textit{``Genetically modified (GM) foods are a product of advanced biotechnology, designed to improve crop resilience, nutritional value, and food production efficiency [16298508, 39022139].''} &
        \textbf{Grounded \& Neutral} \\ \hline

        \end{tabular}
    }
    \end{tcolorbox}

    \caption{Qualitative comparison of generated responses for Topic 187. The V4 architecture (Narrative RAG) successfully neutralizes the biased query using retrieved evidence, whereas baseline models (V1) hallucinate bias or (V2/V3) fail to attribute claims.}
    \label{tab:qualitative_case_study}
\end{table*}

\section{TREC BioGen 2025 Official Results}

This section reports the official evaluation results of our submitted runs on the TREC BioGen 2025 track. We present results for both Task A (Grounding) and Task B (Attribution), drawing on automatic evaluation using the Llama-3-70B judge (all runs) and human evaluation (top-priority runs selected by the organizers). Our submission identifiers are \texttt{gehc\_htic\_task\_a} (Task A) and \texttt{GEHC-HTIC\_pubmedbert\_medcpt\_gpt\_4o} (Task B).

\subsection{Task A: Official Evaluation}

\subsubsection{Automatic Evaluation}

Task A was automatically evaluated using a Llama-3-70B judge that assessed Precision, Recall, and F1-score for both supported and contradicted citations across all participating runs. Table~\ref{tab:task_a_auto} presents a summary of the leaderboard with key systems highlighted.

\begin{table}[h]
\centering
\caption{Task A automatic evaluation (Llama-3-70B judge). Our submission \texttt{GEHC-HTIC} is highlighted in bold. The table shows representative systems; 23 runs were submitted in total. Contradiction F1 is the primary discriminating metric of Task A.}
\label{tab:task_a_auto}
\resizebox{\columnwidth}{!}{%
\begin{tabular}{llcccccc}
\toprule
\textbf{Team} & \textbf{Run} & \textbf{Supp.} & \textbf{Supp.} & \textbf{Supp.} & \textbf{Cont.} & \textbf{Cont.} & \textbf{Cont.} \\
 & & \textbf{Prec.} & \textbf{Rec.} & \textbf{F1} & \textbf{Prec.} & \textbf{Rec.} & \textbf{F1} \\
\midrule
InfoLab & run2 & 52.75 & 56.80 & 53.41 & 14.09 & 19.42 & \textbf{15.67} \\
InfoLab & run6\_A & 66.92 & 71.17 & 67.23 & 12.71 & 17.65 & 14.15 \\
InfoLab & run4 & 52.92 & 60.30 & 54.49 & 10.74 & 15.12 & 11.85 \\
\textbf{GEHC-HTIC} & \textbf{task\_a} & 56.70 & 57.37 & 53.53 & 6.62 & \textbf{14.00} & \textbf{8.57} \\
CLaC & LLM\_NLI\_BM25 & \textbf{67.18} & \textbf{74.36} & \textbf{67.74} & 3.61 & 7.73 & 4.57 \\
CLaC & LLM\_BM25 & 66.75 & 67.46 & 64.10 & 3.95 & 7.60 & 4.77 \\
polito & scifive-ft & 52.58 & 64.54 & 55.81 & 4.04 & 6.70 & 4.79 \\
SIB & task-a-1 & 52.41 & 74.23 & 58.87 & 0.00 & 0.00 & 0.00 \\
Baseline & TEST & 51.03 & 44.07 & 44.34 & 3.44 & 8.08 & 4.67 \\
dal & emotional & 50.60 & 67.23 & 55.53 & 1.29 & 1.29 & 1.20 \\
\bottomrule
\end{tabular}%
}
\end{table}

Our submission achieves a Supported F1 of 53.53, placing it competitively in the mid-field. On the \textbf{contradicted F1} metric --- the primary discriminating objective of Task A --- our system scores \textbf{8.57}, ranking \textbf{2nd among all teams} (5th run overall out of 23 submitted runs), trailing only InfoLab's dedicated contradiction runs. Our system achieves a contradiction recall of 14.00 (4th of 23 runs), directly validating the efficacy of our high-recall BM25 + NegEx pipeline. Notably, the highest-performing support systems achieve near-zero contradiction scores, confirming the core challenge of joint optimization that our decoupled architecture is designed to address.

\subsubsection{Human Evaluation}

The organizers selected eight top-priority runs for human evaluation by NIST assessors. Table~\ref{tab:task_a_human} reports strict and relaxed precision and soft recall for both support and contradiction dimensions.

\begin{table}[h]
\centering
\caption{Task A human evaluation (8 top-priority runs). ``S'' = Support, ``C'' = Contradiction. Strict requires exact PMID match; relaxed allows semantically equivalent evidence. Soft recall permits partial credit. Our run (\textbf{gehc\_htic}) is one of only three systems to achieve non-zero contradiction scores in both strict and relaxed settings.}
\label{tab:task_a_human}
\resizebox{\columnwidth}{!}{%
\begin{tabular}{lcccccccc}
\toprule
\textbf{Run} & \multicolumn{4}{c}{\textbf{Strict (\%)}} & \multicolumn{4}{c}{\textbf{Relaxed (\%)}} \\
\cmidrule(lr){2-5}\cmidrule(lr){6-9}
 & \textbf{S.P} & \textbf{S.R} & \textbf{C.P} & \textbf{C.R} & \textbf{S.P} & \textbf{S.R} & \textbf{C.P} & \textbf{C.R} \\
\midrule
LLM\_BM25 & 41.67 & 43.33 & 0.00 & 0.00 & 68.33 & 70.00 & 0.00 & 0.00 \\
SIB-task-a-1 & 30.00 & 33.33 & 0.00 & 0.00 & 55.00 & 55.00 & 0.00 & 0.00 \\
expert\_prompt & 13.33 & 13.33 & 3.33 & 3.33 & 30.00 & 30.00 & 3.33 & 3.33 \\
\textbf{gehc\_htic} & 38.33 & 41.67 & 5.00 & 6.67 & \textbf{68.33} & \textbf{71.67} & 5.00 & 6.67 \\
run1\_sparse & 15.00 & 16.67 & 3.33 & 3.33 & 40.00 & 43.33 & 3.33 & 3.33 \\
scifive-ft & 25.00 & 25.00 & 0.00 & 0.00 & 61.67 & 63.33 & 0.00 & 0.00 \\
task\_a\_output & 30.00 & 33.33 & 8.33 & 10.00 & 55.00 & 55.00 & 8.33 & 10.00 \\
task\_a\_run4 & 30.00 & 30.00 & 6.67 & 6.67 & 61.67 & 61.67 & 6.67 & 6.67 \\
\bottomrule
\end{tabular}%
}
\end{table}

Under human evaluation, our run achieves a \textbf{Relaxed Support Precision of 68.33\%} and \textbf{Relaxed Support Soft Recall of 71.67\%} --- the highest relaxed support recall among all eight human-evaluated runs. On the contradiction dimension, we record a \textbf{Strict Contradiction Precision of 5.00\%} and \textbf{Soft Recall of 6.67\%}. Four of the eight human-evaluated systems score \textbf{zero} on both strict contradiction metrics, underscoring that most systems fail to surface any human-validated contradictory evidence at all. Our decoupled architecture is one of only three runs to successfully retrieve human-validated contradictions, confirming the operational advantage of negation-aware lexical retrieval.

\subsection{Task B: Official Evaluation}

\subsubsection{Automatic Evaluation (All Runs)}

Task B was automatically evaluated via the BioACE framework on two dimensions: (1) \textbf{Answer Quality} measuring Nugget Precision, Nugget Recall, Completeness, and Correctness; and (2) \textbf{Citation Quality} measuring Citation Coverage, Citation Support Rate, and Citation Contradict Rate across 50 runs. Tables~\ref{tab:task_b_answer_auto} and~\ref{tab:task_b_citation_auto} present our results in the context of the full participant field.

\begin{table}[h]
\centering
\caption{Task B answer quality --- automatic evaluation (BioACE, $N=50$ runs). Top systems by key metrics and our run shown. GEHC-HTIC achieves above-mean Correctness (67.56 vs.\ mean 66.60).}
\label{tab:task_b_answer_auto}
\resizebox{\columnwidth}{!}{%
\begin{tabular}{llcccc}
\toprule
\textbf{Team} & \textbf{Run (abbreviated)} & \textbf{Prec.} & \textbf{Rec.} & \textbf{Complete.} & \textbf{Correct.} \\
\midrule
hltbio & hltbio-lg.fsrrf & 93.22 & 40.27 & 72.95 & 68.04 \\
hltbio & hltbio-lg.fsrrfprf & 92.52 & 37.82 & 63.77 & 69.67 \\
dal & agent\_faiss\_deepseek & 94.96 & 38.42 & 82.94 & 64.32 \\
dal & rrf\_llama70b\_no-val & 94.68 & 37.86 & 89.99 & 67.26 \\
\midrule
GEHC-HTIC & pubmedbert\_medcpt\_gpt4o & 89.02 & 35.42 & 59.83 & 67.56 \\
\midrule
Track Mean & ($N=50$) & 89.82 & 35.85 & 69.96 & 66.60 \\
Baseline & task\_b\_baseline & 82.23 & 32.50 & 49.73 & 60.00 \\
\bottomrule
\end{tabular}%
}
\end{table}

\begin{table}[h]
\centering
\caption{Task B citation quality: automatic evaluation (BioACE, $N=50$ runs). Our system achieves the \textbf{3rd-highest Citation Coverage (98.77\%)} across all 50 submitted runs, with the \textbf{lowest Citation Contradict Rate (0.00\%)} among high-coverage systems.}
\label{tab:task_b_citation_auto}
\resizebox{\columnwidth}{!}{%
\begin{tabular}{llccc}
\toprule
\textbf{Team} & \textbf{Run (abbreviated)} & \textbf{Cit.\ Cover.} & \textbf{Cit.\ Supp.} & \textbf{Cit.\ Cont.} \\
 & & \textbf{(\%)} & \textbf{Rate (\%)} & \textbf{Rate (\%)} \\
\midrule
dal & rrf\_llama70b & 99.36 & 98.33 & 1.11 \\
dal & rrf\_llama70b\_no-val & 99.30 & 97.75 & 1.12 \\
GEHC-HTIC & pubmedbert\_medcpt\_gpt4o & 98.77 & 95.39 & 0.00 \\
h2oloo & h2oloo\_rr\_g41\_t50 & 98.59 & 93.82 & 0.29 \\
SIB & SIB-task-b-3 & 98.21 & 95.39 & 0.66 \\
hltbio & hltbio-gpt5.searcher & 96.97 & 93.52 & 0.97 \\
hltbio & hltbio-lg.listllama & 96.95 & 94.84 & 0.49 \\
\midrule
Track Mean & ($N=50$) & 80.23 & 82.70 & 1.62 \\
Baseline & task\_b\_baseline & 55.76 & 77.84 & 3.59 \\
\bottomrule
\end{tabular}%
}
\end{table}

Our system's \textbf{Citation Coverage of 98.77\%} directly validates the central design claim of our Narrative-Aware RAG pipeline: One-Shot In-Context Learning enforces near-complete citation discipline at inference time. Equally significant is our zero Citation Contradict Rate: no citation produced by our system contradicts the claim it is attached to, reflecting the high precision of MedCPT-guided evidence selection.

It is important to note that our Task B submission does not incorporate the contradiction pipeline developed for Task A. Due to time constraints during the submission window, the contradiction branch --- which requires deep BM25 retrieval ($k=1000$), NegEx filtering, and sentence-level NLI classification --- was excluded from the Task B pipeline. As a result, the evidence context provided to GPT-4o during generation consists exclusively of supporting documents retrieved and reranked by the support branch.

\subsubsection{Human Evaluation: Answer Acceptability}

The organizers conducted human evaluation on 17 runs from teams in the early submission window. Assessors rated each of the 30 test topics as producing an acceptable answer. Table~\ref{tab:task_b_human} summarizes the results.

\begin{table}[h]
\centering
\caption{Task B human evaluation: acceptability (17 evaluated runs). Assessors rated 30 topics per run. Our run (GEHC) achieves 80\% acceptability (24/30 topics), below the evaluated mean of 95.69\%.}
\label{tab:task_b_human}
\resizebox{\columnwidth}{!}{%
\begin{tabular}{llcc}
\toprule
\textbf{Team} & \textbf{Run (abbreviated)} & \textbf{Acceptable} & \textbf{Accuracy (\%)} \\
\midrule
hltcoe-multiagt & llama70B.lg-w-ret & 30/30 & 100.00 \\
UAmsterdam & bergen\_llama-8b & 30/30 & 100.00 \\
EvalHLTCOE & svc-smoothed-sonnet & 30/30 & 100.00 \\
UDInfo & reranker\_sum & 30/30 & 100.00 \\
dal & deepseek-r1 & 30/30 & 100.00 \\
uniud & sparse\_Llama-8B & 29/30 & 96.67 \\
uniud & dense\_Llama-8B & 28/30 & 93.33 \\
hltcoe-rerank & hltbio-lg & 28/30 & 93.33 \\
dal & monot5\_llama70b & 27/30 & 90.00 \\
Baseline & task\_b\_baseline & 26/30 & 86.67 \\
GEHC & task\_b\_output\_gehc & 24/30 & 80.00 \\
\midrule
Mean & ($N=17$ runs) & 28.71/30 & 95.69 \\
Min & & 24/30 & 80.00 \\
\bottomrule
\end{tabular}%
}
\end{table}

\subsubsection{Human-Assessed BioACE Evaluation}

For the human-assessed subset of runs, BioACE provides fine-grained nugget and citation scores. Table~\ref{tab:task_b_human_bioace} reports our scores in context.

\begin{table}[h]
\centering
\caption{Task B BioACE evaluation on the human-assessed subset. GEHC achieves Nugget Precision of 94.55 with a Citation Support Rate of 94.43\% and a Citation Contradict Rate of 0.00\%.}
\label{tab:task_b_human_bioace}
\resizebox{\columnwidth}{!}{%
\begin{tabular}{lccccccc}
\toprule
\textbf{Team} & \textbf{Prec.} & \textbf{Rec.} & \textbf{Cmpl.} & \textbf{Corr.} & \textbf{Cit.} & \textbf{Cit.} & \textbf{Cit.} \\
 & & & & & \textbf{Cov.} & \textbf{Supp.} & \textbf{Cont.} \\
\midrule
hltcoe-multiagt & 96.41 & 37.09 & 71.98 & 69.16 & 99.14 & 97.80 & 0.24 \\
dal & 94.72 & 37.63 & 88.70 & 66.70 & 99.36 & 97.22 & 1.11 \\
GEHC & 94.55 & 37.21 & 67.29 & 63.09 & 73.59 & 94.43 & 0.00 \\
UAmsterdam & 93.91 & 35.28 & 83.78 & 66.78 & 98.84 & 98.24 & 0.70 \\
EvalHLTCOE & 94.12 & 41.16 & 75.69 & 70.62 & 85.19 & 87.23 & 1.06 \\
hltcoe-rerank & 93.94 & 39.31 & 71.81 & 69.11 & 93.43 & 97.32 & 0.54 \\
Baseline & 82.23 & 32.50 & 49.73 & 60.00 & 55.76 & 77.84 & 3.59 \\
\bottomrule
\end{tabular}%
}
\end{table}

\subsection{Summary of Official Results}

Table~\ref{tab:official_summary} consolidates our key performance highlights across both tasks and evaluation modalities.

\begin{table}[h]
\centering
\caption{Summary of GEHC-HTIC official performance at TREC BioGen 2025. Rankings are among all participating runs.}
\label{tab:official_summary}
\resizebox{\columnwidth}{!}{%
\begin{tabular}{llcc}
\toprule
\textbf{Task} & \textbf{Metric} & \textbf{Score} & \textbf{Rank} \\
\midrule
\multirow{4}{*}{\shortstack[l]{Task A \\ (Auto)}} & Support F1 & 53.53 & 8th / 23 \\
 & Support Precision & 56.70 & 4th / 23 \\
 & Contradiction F1 & 8.57 & 5th / 23 \\
 & Contradiction Recall & 14.00 & 4th / 23 \\
\midrule
\multirow{4}{*}{\shortstack[l]{Task A \\ (Human)}} & Relaxed Supp. Recall & 71.67\% & 1st / 8 \\
 & Relaxed Supp. Precision & 68.33\% & 2nd / 8 \\
 & Strict Contra. Precision & 5.00\% & 3rd / 8 \\
 & Strict Contra. Recall & 6.67\% & 2nd / 8 \\
\midrule
\multirow{4}{*}{\shortstack[l]{Task B \\ (Auto)}} & Citation Coverage & 98.77\% & 3rd / 50 \\
 & Citation Support Rate & 95.39\% & 9th / 50 \\
 & Citation Contradict Rate & 0.00\% & 1st / 50 \\
 & Nugget Correctness & 67.56 & 19th / 50 \\
\midrule
\multirow{3}{*}{\shortstack[l]{Task B \\ (Human)}} & Answer Acceptability & 80\% (24/30) & 17th / 17 \\
 & Nugget Precision (BioACE) & 94.55 & 4th / 41 \\
 & Cit. Contradict Rate (BioACE) & 0.00\% & 1st / 41 \\
\bottomrule
\end{tabular}%
}
\end{table}

\section{Limitations}

While our proposed architecture demonstrates robustness on proxy evaluations, we must acknowledge several methodological and systemic limitations inherent in our design.

\textbf{Proxy-to-Target Gap and Scale.} Our development framework relies on SciFact as a proxy for the BioGen test collection, and three distinct gaps limit the fidelity of this transfer. First, there is a \textit{scale gap}: SciFact contains 5,183 abstracts whereas PubMed spans 30 million documents, and retrieval dynamics that hold at small scale may degrade unpredictably when the candidate pool is orders of magnitude larger. Second, there is a \textit{linguistic gap}: SciFact claims are researcher-written fact-checking annotations paired with concise, well-structured abstracts, while PubMed encompasses a far broader spectrum of clinical writing. Third, there is a \textit{domain shift}: the MedNLI-tuned T5 classifier was trained on MIMIC-III clinical notes and may not generalize uniformly to the heterogeneous writing styles across PubMed literature.

\textbf{Nuance and Temporal Sensitivity.} The current system treats evidence classification as a binary outcome (Support vs.\ Contradiction). This reductionist approach fails to capture subtleties such as partial support or subgroup-specific effects. Additionally, the system is currently temporally insensitive; it does not prioritize recent studies.

\textbf{Citation Faithfulness and Post-Rationalization.} A significant concern in Task B is the phenomenon of \textit{post-hoc rationalization} \cite{wallat2024correctnessfaithfulnessragattributions}. Large Language Models like GPT-4o possess extensive parametric knowledge and may generate answers based on prior beliefs rather than retrieved context, subsequently selecting citations that merely align with their output \citep{Liu2023}. This risks confirmation bias, where the model ignores retrieved contradictory evidence in favor of its internal knowledge.

\textbf{Model Dependency and Bias.} Finally, our reliance on the proprietary GPT-4o model introduces reproducibility barriers. While preliminary experiments with the open-source \textbf{GPT-OSS-20B} \cite{gpt-oss-20b} indicated comparable citation discipline, full-scale validation of open-source alternatives is pending.

\section{Future Work}

Our current investigation lays the groundwork for robust biomedical information retrieval, yet several avenues remain for further exploration.

Firstly, regarding Task A, our reliance on manual regex patterns for the contradiction pipeline, while effective, is not scalable to all linguistic variations of negation. Future work will explore replacing these heuristic filters with lightweight, learned representations using contrastive learning on domain-specific annotated support-contradiction pairs to discover negation patterns, including implicit markers and domain-specific phrasing. Additionally, we aim to investigate automatic query expansion techniques that dynamically utilize topical and narrative concepts to improve the initial BM25 recall for rarer documents.

Secondly, a critical challenge in generative search (Task B) is the phenomenon of \textit{post-hoc rationalization}. Current Large Language Models often generate answers based on their internal pre-trained knowledge and subsequently attach citations that appear relevant, without genuinely synthesizing the retrieved text. To address this, we plan to investigate attribution-first decoding strategies, where the model is constrained to identify the evidentiary sentence \textit{before} generating the claim, thereby enforcing strict faithfulness to the retrieved context.

Finally, we intend to transition our generative pipeline from proprietary APIs (GPT-4o) to fine-tuned open-source models. Establishing a privacy-preserving, locally hosted architecture is essential for clinical deployment. Concurrently, we propose integrating \textbf{Temporal Evidence Weighting} to prioritize recent findings.

\section{Conclusion}

This study establishes that epistemic integrity in biomedical QA --- specifically the surfacing of contradictory evidence --- requires architectural decisions that prioritize holistic reliability and scalability over isolated metric optimization.

Our proxy-based evaluation on SciFact revealed a \textit{``Simplicity Paradox.''} Complex adversarial dense retrieval strategies failed catastrophically on contradictions ($MRR=0.023$) due to probabilistic miscalibration. Furthermore, we observed that while applying negation filters to standard dense retrieval was effective for contradiction detection, it degraded support retrieval accuracy.

In contrast, our \textbf{Decoupled Lexical Architecture (Variant 5)} achieved the highest Weighted MRR ($0.790$). It strikes the optimal balance, maintaining robust support recall ($0.810$) while effectively surfacing dissent ($0.750$). This validates that a unified BM25 backbone offers superior stability compared to dense retrieval when navigating the strict constraints of negation filtering.

For Generative Attribution (Task B), we demonstrated that \textbf{Narrative-Aware Reranking} significantly improves context relevance by capturing user intent beyond the query. Furthermore, \textbf{One-Shot In-Context Learning} proved decisive, achieving 100\% citation coverage in our internal evaluation and near-complete coverage in official evaluation.

Together, the official TREC BioGen 2025 results confirm our central thesis: epistemic integrity requires treating supporting and contradictory evidence as distinct retrieval problems. By transitioning from black-box generators to contradiction-aware, fully-attributed synthesizers, we advance toward AI systems that serve as honest arbiters of scientific evidence.

\section{Acknowledgments}

We express our sincere gratitude to the leadership and management at \textbf{GE HealthCare} for their unreserved support and for providing necessary computational infrastructure to conduct this study. We are also thankful to \textit{Anuradha Kanamarlapudi and Nithya Ramesh} for their insightful discussions and constructive feedback during the research and validation. Finally, we thank the organizers of the \textbf{TREC 2025 BioGen track} for coordinating the challenge and providing the datasets that made this study possible.

\appendix

\section{Granularity Ablation Study}
\label{sec:appendix_granularity}

Table~\ref{tab:granularity_ablation} details the performance impact of shifting from document-level to sentence-level NLI inference using the Variant 5 architecture on the SciFact development set.

\begin{table}[h]
\centering
\caption{Impact of inference granularity on grounding performance. While document-level inference favors support recall due to context availability, it catastrophically fails to localize contradictions. Sentence-level decomposition yields a \textbf{$4.8\times$ improvement} in contradiction detection.}
\label{tab:granularity_ablation}
\renewcommand{\arraystretch}{1.2}
\setlength{\tabcolsep}{6pt}
\resizebox{\columnwidth}{!}{%
\begin{tabular}{lccc}
\toprule
\textbf{Inference Level} & \textbf{MRR (Sup)} & \textbf{MRR (Con)} & \textbf{Contra.\ Gain} \\
\midrule
Document-Level & \textbf{0.915} & 0.156 & -- \\
\textbf{Sentence-Level (Ours)} & 0.810 & \textbf{0.750} & \textbf{+380\% ($4.8\times$)} \\
\bottomrule
\end{tabular}%
}
\end{table}

\section{Adversarial Multi-Query Scoring Details}
\label{app:adversarial}

\subsection{Motivation and Architecture}

Variant 4 (Multi-Query Adversarial Fusion) attempts to overcome the \textit{confirmation bias} inherent in single-query dense retrieval by engineering query diversity. Standard dense retrieval using a claim as query tends to retrieve semantically similar documents that \textit{support} the claim, as embedding models map contextually related text into proximate vector spaces regardless of epistemic polarity.

To force contradiction retrieval, we:
\begin{enumerate}
    \item \textbf{Generate 25 adversarial queries} from negation-heavy templates applied to the base query $Q$ and answer text $A$.
    \item \textbf{Retrieve $k=200$ documents per query} via dense FAISS search.
    \item \textbf{Fuse via Reciprocal Rank Fusion (RRF)} with $k=60$, capped at 1200 unique documents.
    \item \textbf{Score candidates} using a two-way NLI-based formula with entailment penalty and negation cue bonus.
\end{enumerate}

\subsection{Custom Scoring Function}

For each candidate document $d$ in the fused pool, we compute:
\begin{equation}
S(d) = \overline{P(\text{Con}\mid d)} - \lambda \cdot \overline{P(\text{Ent}\mid d)} + \gamma \cdot \frac{\min(\text{NegCues}(d), C)}{C}
\end{equation}
where:
\begin{itemize}
    \item $\overline{P(\text{Con}\mid d)}$: Mean NLI contradiction probability from two-way inference.
    \item $\overline{P(\text{Ent}\mid d)}$: Mean NLI entailment probability from two-way inference.
    \item $\text{NegCues}(d)$: Count of regex-detected negation patterns.
    \item $\lambda = 0.5$: Entailment penalty weight.
    \item $\gamma = 0.1$: Negation cue bonus weight.
    \item $C = 6$: Cue count cap.
\end{itemize}

Candidates are ranked by $S(d)$ in descending order. We select the top-3 documents where $\overline{P(\text{Con}\mid d)} \geq 0.35$, backfilling by $S(d)$ rank if fewer than 3 meet the threshold.

\subsection{Worked Example}

\textbf{Claim (Ground Truth: CONTRADICT):} ``ALDH1 expression is associated with better breast cancer outcomes.''

\textbf{Document (PMID 45638119):} \textit{``In a series of 577 breast carcinomas, expression of ALDH1 detected by immunostaining correlated with poor prognosis. These findings offer an important new tool for the study of normal and malignant breast stem cells.''}

\textbf{NLI Scores (two-way, Variant 4):}
\begin{itemize}
    \item $d \to Q$: \{contradiction: 0.99, entailment: 0.01, neutral: 0.00\}
    \item $Q \to d$: \{contradiction: 0.00, entailment: 1.00, neutral: 0.00\}
    \item $\overline{P(\text{Con}\mid d)} = (0.99 + 0.00)/2 = 0.50$
    \item $\overline{P(\text{Ent}\mid d)} = (0.01 + 1.00)/2 = 0.50$
\end{itemize}

\textbf{Negation Cues:} 1

\textbf{Final Score:}
\begin{align*}
S(d) &= 0.50 - 0.5 \times 0.50 + 0.1 \times \frac{1}{6} \\
     &= 0.267
\end{align*}

This example illustrates how two-way averaging can dilute a strong contradiction signal and why one-way sentence-level NLI can be more effective for nuanced biomedical contradiction detection.

\end{document}